\documentclass[aps,prb,twocolumn,superscriptaddress,showpacs,10pt]{revtex4-1}
\usepackage{graphicx}
\usepackage{calc}
\usepackage{bm}
\usepackage{color}

\begin{document}

\title{Current-induced magnetization dynamics at the edge of a two-dimensional electron system with strong spin-orbit coupling}

\author{A.~Kononov}
\affiliation{Institute of Solid State Physics RAS, 142432 Chernogolovka, Russia}
\affiliation{Moscow Institute of Physics and Technology, Institutsky per. 9, Dolgoprudny, 141700 Russia}
\author{S.V.~Egorov}
\affiliation{Institute of Solid State Physics RAS, 142432 Chernogolovka, Russia}
\author{G.~Biasiol}
\affiliation{IOM CNR, Laboratorio TASC, 34149 Trieste, Italy}
\author{L.~Sorba}
\affiliation{NEST, Istituto Nanoscienze-CNR and Scuola Normale Superiore, 56127 Pisa, Italy}
\author{E.V.~Deviatov}
\affiliation{Institute of Solid State Physics RAS, 142432 Chernogolovka, Russia}
\affiliation{Moscow Institute of Physics and Technology, Institutsky per. 9, Dolgoprudny, 141700 Russia}

\date{\today}

\begin{abstract}
We experimentally investigate electron transport through the interface between a permalloy ferromagnet and the edge of a two-dimensional electron system with strong Rashba-type spin-orbit coupling. We observe strongly non-linear transport around zero bias at millikelvin temperatures. The observed nonlinearity is fully suppressed above some critical values of temperature, magnetic field, and current through the interface. We interpret this behavior as the result of spin accumulation at the interface and its current-induced absorption as a magnetization torque.
\end{abstract}


\maketitle

\section{Introduction}

Recently, there is a strong interest in semiconductor hybrid structures, which consist of a metal and a low-dimensional semiconductor structure with strong spin-orbit (SO) coupling. The general interest is devoted to the modification of transport in the low-dimensional  structure caused by the proximity with a metal, which is characterized by a macroscopic order parameter. In the case of a superconducting metal, the interest is mostly stimulated by the search for Majorana fermions.~\cite{reviews}

As a superconductor, also a ferromagnet (F) is characterized by a macroscopic order parameter. In the case of a ferromagnet, there is the possibility of injection and detection of spin-polarized electrons. This is important, e.g., for investigations of the spin-Hall effect, which manifests itself as finit spin accumulation at the sample edges, generated by an electric field in a low-dimensional system with strong SO coupling~\cite{spin_hall}. The existence of the spin-Hall effect~\cite{kato,wunderlich} was firstly confirmed in transport investigations of thin metallic films~\cite{tinkham,saitoh} and much later in optical experiments~\cite{werake,yang} in semiconductors. 

On the other hand, a more general problem can be formulated: the mutual influence of two systems at the interface between them. In the case of a superconducting metal close to a two-dimensional electron gas (2DEG), Andreev reflection is suppressed because the strong SO coupling affects pairing near the interface~\cite{nbinas,inoue}. In the case of a ferromagnet, spin-dependent transport through the F-2DEG interface defines the current-induced magnetization dynamics in a ferromagnetic contact~\cite{halperin06}, e.g.~magnetization torque~\cite{torque}. The latter effect was mostly investigated in  multilayer systems~\cite{miron,myers,chernyshov}, which consist of a set of normal and ferromagnetic layers. A 2DEG realized in a semiconductor quantum well, differs significantly from a thin metallic film. In particular, a 2DEG edge is well-known to exhibit a very specific one-dimensional behavior both in quantizing~\cite{buttiker} and in zero~\cite{shklovskii} magnetic fields. Thus, it is quite reasonable to study spin transport in a F-2DEG planar device located at the edge of a 2DEG with strong Rashba-type SO coupling.

Here, we experimentally investigate electron transport through the interface between a permalloy ferromagnet and the edge of a two-dimensional electron system with strong Rashba-type spin-orbit coupling. We observe strongly non-linear transport around zero bias at millikelvin temperatures. The observed nonlinearity is fully suppressed above some critical values of temperature, magnetic field, and current through the interface. We interpret this behavior as a result of spin accumulation at the interface and its current-induced absorption as a magnetization torque. 

\section{Samples and technique}

\begin{figure}
\includegraphics[width=\columnwidth]{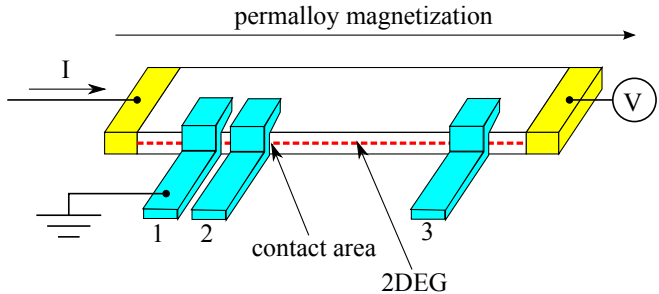}
\caption{(Color online) Sketch of the sample with electrical connections (not in scale). The 100~$\mu$m wide mesa has two Au Ohmic contacts (yellow). Three ferromagnetic $Fe_{20}Ni_{80}$ permalloy stripes (blue, denoted by numbers) are placed to overlap the mesa step. In every overlap region, a planar F-2DEG junction is formed  between the ferromagnetic  film and the 2DEG edge (denoted by a dashed line). The width of each junction is equal to 20~$\mu$m. The junctions 1 and 2 are separated by 2$\mu$m distance, while junction 3 is shifted by 400~$\mu$m  along the mesa edge. We study electron transport across one particular F-2DEG junction in a three-point configuration: the corresponding ferromagnetic electrode is grounded (no.~1 in the figure), others are disconnected; a current is applied between it and one of the Au Ohmic contacts; the other Au contact traces the 2DEG potential.}
\label{sample}
\end{figure}

Our samples are grown by solid source molecular beam epitaxy on semi-insulating GaAs (100)  substrates. The active layer is composed of a 20-nm thick $In_{0.75}Ga_{0.25}As$ quantum well sandwiched between a lower 50-nm thick and an upper 120-nm thick $In_{0.75}Al_{0.25}As$ barrier. Details on the growth parameters can be found elsewhere~\cite{biasiol05,biasiol08}. A two dimensional electron gas, confined in a narrow asymmetric $In_{0.75}Ga_{0.25}As$ quantum well, is characterized by strong Rashba-type SO coupling~\cite{holmes,inas}. For our samples, the 2DEG mobility at 4K is about $5 \cdot 10^{5}  $cm$^{2}$/Vs  and the carrier density is   $4.1 \cdot 10^{11}  $cm$^{-2}$, as obtained from standard magnetoresistance measurements.

A sample sketch is presented in Fig.~\ref{sample}. A 200 nm high mesa  is formed by wet chemical etching. In our $In_{0.75}Ga_{0.25}As$ structure, a high quality contact to a 2DEG edge can be realized by evaporation of a metal over the mesa edge, without annealing procedure~\cite{inas,nbinas}. We fabricate two Ohmic contacts to the 2DEG by thermal evaporation of 100~nm Au (with few nm Ni to improve adhesion). These Ohmic contacts are characterized by a constant ($\approx 1 k\Omega$) resistance. In addition, we use rf sputtering to deposit 50~nm thick ferromagnetic $Fe_{20}Ni_{80}$ permalloy stripes to overlap the  mesa edge. The initial magnetization of the permalloy  is oriented along the mesa edge, see Fig.~\ref{sample}. The stripes are formed by lift-off technique, and the surface is mildly cleaned by Ar plasma before sputtering. To avoid any 2DEG  degradation, the sample is not  heated during the sputtering process.

A planar F-2DEG junction is formed  between the ferromagnetic electrode and the 2DEG at the mesa edge. We study electron transport across one particular F-2DEG junction in a three-point configuration: a current is applied between one of the Au Ohmic contacts and a ferromagnetic electrode which is grounded (contact 1 in Fig.~\ref{sample}) while the other Au contact measures the 2DEG potential. To obtain $dV/dI(V)$ characteristics, we sweep the dc current through the interface from -5~$\mu$A to +5~$\mu$A.  This dc current is modulated by a low (0.85~nA) ac (110~Hz) component. We measure both the dc ($V$) and ac ($\sim dV/dI$) components of the 2DEG potential by using a dc voltmeter and a lock-in amplifier, respectively. We have checked, that the lock-in signal is independent of the modulation frequency in the range 50~Hz -- 300~Hz. This range is defined by applied ac filters. Because of the relatively low in-plane  2DEG resistance (about 100 $\Omega$ at present 2DEG concentration and mobility), and the low resistance of the metallic permalloy electrode, the measured $dV/dI(V)$ curves reflect the behavior of the F-2DEG interface. To extract features specific to the SO coupling, the measurements were performed at a temperature of 30~mK. Similar results were obtained from different samples in several cooling cycles.

\section{Single F-2DEG junctions}
\subsection{Results}

\begin{figure}
\includegraphics[width=\columnwidth]{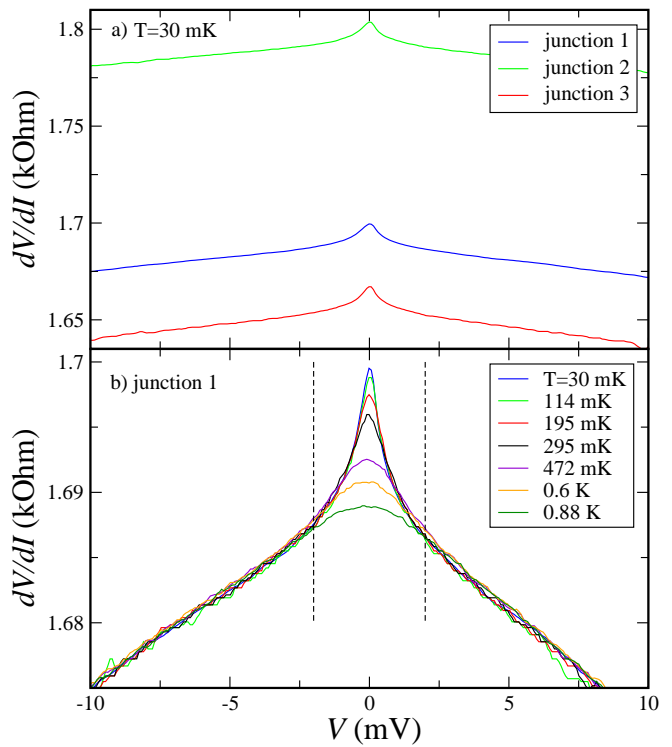}
\caption{(Color online) (a) Differential resistance $dV/dI$ of a single F-2DEG junction as a function of the dc voltage drop $V$ across the  junction. The curves are denoted by the junction numbers, see Fig.~\protect\ref{sample}. Each curve demonstrates a well developed non-linear behavior at low bias. The curves from three F-2DEG junctions only differ by a constant and bias-independent offset. (b) Evolution of  the $dV/dI(V)$ curve with temperature. The non-linear region around zero bias exists only at low temperatures and disappears completely at 0.88~K. On the other hand, the linear branches of the $dV/dI(V)$ curves are invariant in this temperature range.}
\label{IVterm}
\end{figure}

Examples of  $dV/dI(V)$ characteristics are presented in Fig.~\ref{IVterm}(a) for three different F-2DEG junctions, depicted in Fig.~\ref{sample}. All three experimental curves in Fig.~\ref{IVterm}(a) look quite similar: they  only differ by a constant and bias-independent offset. The offset absolute value  does  not correlate with the junction position along the mesa edge: the measured resistance is  maximum for the junction 2, which is not the closest one to the current or voltage Ohmic contacts. This is another experimental verification that the measured resistance is a  characteristics of one particular F-2DEG interface, see also Section~\ref{double}.

Each curve in Fig.~\ref{IVterm}(a) demonstrates strongly non-linear behavior, which is shown in detail in Fig.~\ref{IVterm}(b) for the junction 1. The curve is slightly asymmetric with respect to voltage and is characterized by a strictly linear dependence of $dV/dI(V)$, except in the narrow region around zero bias. A temperature increase suppresses the zero-bias non-linearity. The non-linearity  disappears completely at 0.88~K. In contrast, the linear branches of the $dV/dI(V)$ curve are invariant in this temperature range. 

\begin{figure}
\includegraphics[width=\columnwidth]{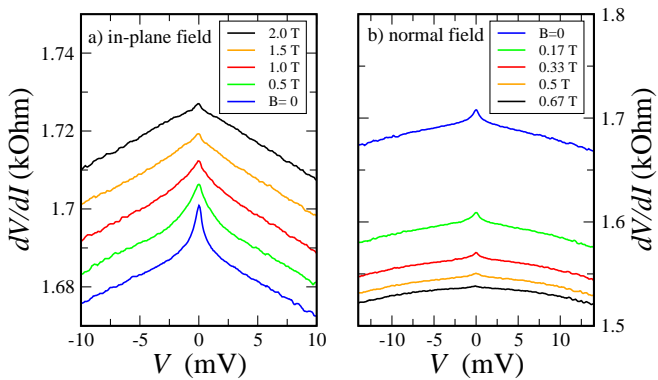}
\caption{(Color online) Suppression of the zero-bias nonlinearity by (a) in-plane or (b) perpendicular magnetic field at $T=30$~mK. The full suppression occurs at $B=2$~T for the in-plane oriented magnetic field while it occurs at much lower field ($B=0.67$~T) for the perpendicular field orientation. The linear branches of the $dV/dI(V)$ curves demonstrate positive (a) or negative (b) bias-independent magnetoresistance.}
\label{IVmag}
\end{figure}
  
Similarly to temperature, the zero-bias nonlinearity in $dV/dI$ can be suppressed by a magnetic field, as shown in Fig.~\ref{IVmag}. The full suppression occurs at quite high $B=2$~T for an in-plane oriented field while it occurs at much lower field ($B=0.67$~T) for the perpendicular field orientation. The linear branches of the $dV/dI(V)$ curves demonstrate bias-independent magnetoresistance, which is positive (a) or negative (b), depending on the magnetic field orientation.

\subsection{Discussion}

Let us start the discussion with the  $dV/dI(V)$ curves with suppressed zero-bias nonlinearity, i.e. from Fig.~\ref{IVterm}(b) at higher temperature ($0.88K$) or from Fig.~\ref{IVmag}(a) above 1.5~T. The linear dependence of $dV/dI(V)$, outside the zero bias region as well as the clear asymmetry of the curve, differs significantly from usual Ohmic behavior with constant $dV/dI(V)$. This behaviour indicates the presence of a (narrow) potential barrier at the interface between the ferromagnet and the 2DEG, e.g.~due to depletion at the 2DEG edge~\cite{shklovskii}. From the constant slope of the linear branches in Figs.~\ref{IVterm} and \ref{IVmag} we can conclude that in our experiment the barrier is roughly independent of temperature and  magnetic field. Apart from the barrier,  the junction resistance is affected by single-particle scattering due to 2DEG disorder in the vicinity of the interface. The  disorder is responsible  for different offset values observed for different junctions as shown  in Fig.~\ref{IVterm}(a). Both  barrier and  disorder define a single-particle transmission of the interface which we can estimate to be $T\approx 0.1$ from the junction resistance and  width. This $T$ value is slightly different (below 10\%) for different junctions, which is in agreement with the disorder variation observed in Ref.~\onlinecite{biasiol08} for similar samples.  

\begin{figure}
\includegraphics[width=0.85\columnwidth]{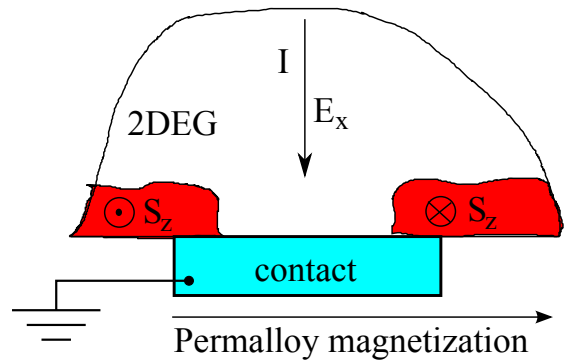}
\caption{(Color online) Top-view of the 2DEG region near the ferromagnetic contact (see Ref.~\protect\onlinecite{halperin04} for details) with zero magnetic field for the case of a strong Rashba SO coupling. The electric current $I$ is flowing through the 2DEG to the F-2DEG interface. The corresponding electric field $E_x$ creates a non-zero, out-of-plane spin polarization $S_z$ around the  junction corners (red regions, not in scale), because of a spin-Hall effect. The permalloy film magnetization is oriented within the 2DEG plane. }
\label{spin}
\end{figure}

However, both disorder and potential barrier are junction-specific values, so they  cannot be responsible for the quite universal zero-bias nonlinearity in $dV/dI$ at low temperature. On the other hand, characteristic values of the non-linearity suppression ($T\approx$~1~K and $B\approx$~1.5-2~T in-plane field) are well known for the 2DEG in our $In_{0.75}Ga_{0.25}As$ quantum well: the spin-orbit splitting $\Delta_{SO}$ is about 0.1~meV in zero magnetic field~\cite{nbinas,holmes},  while Zeeman splitting exceeds~\cite{nbinas,holmes} $\Delta_{SO}$  for in-plane $B > 1.5$~T.   For this reason and because the observed non-linearity is quite universal for different junctions, it seems to be reasonable to connect the observed nonlinearity with spin effects due to the Rashba SO coupling.

The spin effects are expected to be quite sophisticated in the vicinity of the interface~\cite{halperin04}. At zero magnetic field, the electric field $E_x$, which originates from flowing current, is expected to cause a non-zero spin current $j^z_y$ in a clean, infinite and homogeneous 2DEG~\cite{sinova}, which is well known as a spin-Hall effect. However, the spin current is not measurable directly, so its
physical meaning is obscure~\cite{halperin04}. A more meaningful quantity is spin polarization (spin accumulation) rather
than a spin current. Calculations that included scattering~\cite{dirty} resulted in $j^z_y=0$, and $j^z_y=0$ has been also proven directly even in absence of scattering~\cite{prb}. Despite this fact, spin polarization $S_z$ near the edges even in absence of spin current has been found in a number of theoretical papers, see, e.g., Refs.~\onlinecite{rashba,khaetskii}.  Thus, out-of-plane spin polarization $S_z$ is accumulated around the  corners~\cite{halperin04}, see Fig.~\ref{spin}. Since the permalloy film has in-plane magnetization, transport of out-of-plane polarized electrons to the contact is difficult because the necessary absorption of a polarization component perpendicular to the permalloy magnetization. The junction width is effectively diminished, which gives rise to the increased differential resistance $dV/dI$ around zero bias as shown in Fig.~\ref{IVterm}.

When we increase the current through the interface, this out-of-plane spin polarization can be transferred to the permalloy magnetization as a magnetization torque~\cite{torque}. This restores the contact effective width, so the differential resistance is diminished exactly to the same values as obtained by a temperature increase as shown in Fig.~\ref{IVterm}(b). The current-induced polarization absorption is characterized~\cite{torque} by some critical current value, which  can be estimated from Fig.~\ref{IVterm}(b) as $\approx 1 \mu$A . This relatively low~\cite{halperin06} value originates from the specific geometry: the planar junction is formed by a thin permalloy film at vertical mesa edge. If we consider the finite 2DEG thickness, we obtain quite reasonable~\cite{torque} critical current density of $10^{4}-10^{5}$~A/cm$^2$.

The above picture is essentially based on the presence of a strong Rashba-type spin-orbit coupling in the 2DEG. If the temperature exceeds the value of the spin-orbit splitting $\Delta_{SO} \approx 0.1$~meV, all the effects of the spin polarization  disappear in the 2DEG, and the interface resistance is diminished, as we do observe in Fig.~\ref{IVterm}(b) at $T\approx 1$~K. The nonlinearity observed in Fig.~\ref{IVterm} is indeed induced by the current, because the nonlinearity bias range  in Fig.~\ref{IVterm}(b) (about 4~mV) exceeds significantly the characteristic suppression temperature $T\approx 1$~K. An in-plane magnetic field has a similar effect: when the Zeeman splitting  exceeds $\Delta_{SO}$ at $B\approx 1.5$~T, the spin-Hall effect disappears, and the interface resistance is diminished, as can be seen in Fig.~\ref{IVmag}(a). A positive, bias-independent magnetoresistance of the linear branches of the $dV/dI(V)$ curves reflects the spin-polarization of the 2DEG~\cite{dolgop} and is not sensitive to spin-orbit effects.

The primary effect of a perpendicular magnetic field is different. It easily aligns already at lower magnetic field values  ($B\approx 0.67$~T) the magnetization of the soft permalloy ferromagnet to the field direction, i.e. in the $S_z$ direction in this case. The transport through the F-2DEG interface does not require the perpendicular magnetization component absorption. Thus, the junction resistance is reduced, as we see in Fig.~\ref{IVmag}(b). Negative magnetoresistance of the linear branches of $dV/dI(V)$ curves is defined by the 2DEG orbital effects in a perpendicular field.

\section{Double F-2DEG-F junctions}\label{double}

\begin{figure}
\includegraphics[width=\columnwidth]{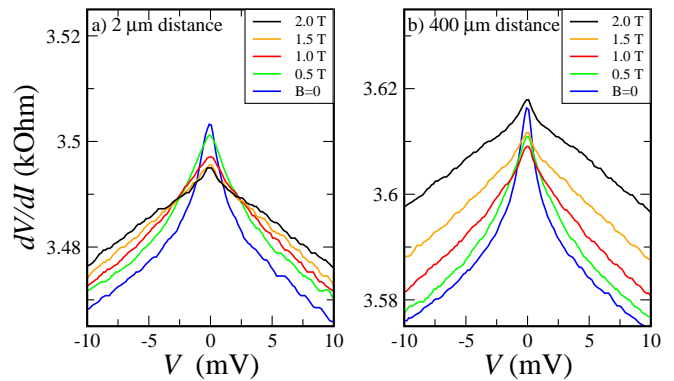}
\caption{(Color online) Two-point differential resistance $dV/dI$ between two ferromagnetic leads (F-2DEG-F junction). (a) For the shorter distance between the leads (2~$\mu$m), the $dV/dI(V)$ curve is exactly the sum of the $dV/dI(V)$ characteristics of two F-2DEG junctions (at zero field). (b) For larger distance between the leads (400~$\mu$m), the $dV/dI(V)$ curve contains a noticeable (about 100~$\Omega$) 2DEG resistance between the leads, compare the corresponding curves in Fig.~\ref{IVterm}(a). In-plane magnetic field suppresses the zero-bias nonlinearity equally in both cases, demonstrating that it is determined by spin effects at the interface.}
\label{IV2p}
\end{figure}

The above arguments are supported by the magnetic field behavior of F-2DEG-F junctions. We measured the bias-dependent differential resistance of F-2DEG-F junctions in a two-point configuration, by grounding one ferromagnetic stripe in Fig.~\ref{sample} and using another to apply a current and to measure a voltage drop simultaneously.

First we will show that the non-linearity is determined by the interface, see Fig~\ref{IV2p}. The contacts 1 and 2 are separated by a distance of 2~$\mu$m, which is below the 10~$\mu$m mean free path in the 2DEG. It is thus not surprising that the $dV/dI(V)$ curve in Fig~\ref{IV2p}(a) at $B=0$ is exactly the sum of the two corresponding $dV/dI(V)$ characteristics from Fig.~\ref{IVterm}(a).  The positive magnetoresistance is very weak in this case, because the ballistic transport is less sensitive to the 2DEG between the two ferromagnetic contacts. For the long (400~$\mu$m) F-2DEG-F junction, see Fig~\ref{IV2p}(b), the $dV/dI(V)$ curve contains a noticeable (about 100~$\Omega$) 2DEG resistance even in zero magnetic field (compare the corresponding values in Figs.~\ref{IVterm}(a) and \ref{IV2p}(b)). In this case the positive magnetoresistance is practically restored. 

However, in both cases, the  zero-bias nonlinearity suppression by the in-plane magnetic field is exactly the same as in the case of a single junction. We can thus conclude that the non-linearity is determined by the  interface. 

\begin{figure}
\includegraphics[width=\columnwidth]{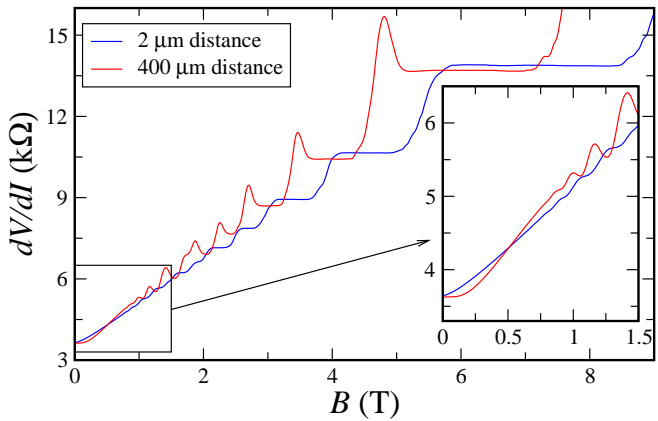}
\caption{(Color online) Two-point magnetoresistance $dV/dI$ between two ferromagnetic leads (F-2DEG-F) at zero bias in perpendicular field. For large (500~$\mu$m) distance between the leads (red line), the $dV/dI (B)$ curve is the sum of the $xx$ and $xy$ magnetoresistance components, as expected for a two-point measurements. However, for the short (2~$\mu$m) distance between the leads (blue line), the $dV/dI (B)$ demonstrates a clear quantum Hall behavior. The inset shows that this behavior occurs even at very low magnetic fields.}
\label{R2p}
\end{figure}

Next, ballistic, spin-dependent transport in the vicinity (about 2~$\mu$m) of a junction is demonstrated in Fig.~\ref{R2p}. For larger distance (400~$\mu$m) between the leads (red line), the $dV/dI (B)$ curve is  a sum of $xx$ and $xy$ magnetoresistance components, as expected for a quantum Hall two-point measurements. However, the $dV/dI (B)$ demonstrates a clear Hall (xy) behavior for the short (2~$\mu$m) distance between the leads (blue line) already at very low magnetic fields, see inset to Fig.~\ref{R2p}.

In a quantizing magnetic field, the bulk spectrum of a 2DEG is a Landau ladder with additional Zeeman (spin) sub-splitting. Current-carrying edge states   at the sample edge~\cite{buttiker} are therefore characterized by the out-of-plane spin projection. A perpendicular magnetic field easily aligns the magnetization of the soft permalloy ferromagnet to the field direction. In the case of two-point measurements, (out-of-plane) spin-polarized electrons are injected into the edge state with the same spin projection. For the short F-2DEG-F junction, electrons travel along the edge states and are absorbed in the other ferromagnetic electrode with the same spin projection. In the case of a long junction, charge redistribution takes place between the edge states, which is accompanied by a spin-flip. Thus,  only part of the electrons can be absorbed at the end. Therefore, for the short F-2DEG-F junction, we have a perfect quantum Hall behavior even for the two-point measurements, while in the larger junction we have the usual sum of the xx and xy resistance components. This behavior demonstrates  ballistic, spin-dependent transport in the close vicinity of the F-2DEG interface.

\section{Conclusion}
We experimentally investigate electron transport through the interface between a permalloy ferromagnet and the edge of a two-dimensional electron system with strong Rashba-type spin-orbit coupling. We observe strongly non-linear transport around zero bias at millikelvin temperatures. The observed nonlinearity is fully suppressed above some critical values of temperature, magnetic field, and current through the interface. We interpret this behavior as a result of spin accumulation at the interface and its current-induced absorption as a magnetization torque. 
 
\acknowledgments

We wish to thank V.T.~Dolgopolov for fruitful discussions, and Stefan Heun for critical reading of the manuscript.  We gratefully acknowledge financial support by the RFBR (project No.~13-02-00065) and RAS.

\end{document}